\documentclass[aps,prb,twocolumn,superscriptaddress,floatfix,showpacs]{revtex4-1}
\usepackage{tikz}
\usepackage{amsmath}
\usepackage{makeidx,amssymb,amsmath,amsthm,graphicx}
\usepackage[toc,page]{appendix}

\begin{document}

%\preprint{APS/SmRh2Si2}

\title{Bulk properties of single crystals of the valence-unstable compound SmRh$_2$Si$_2$}

\author{K.Kliemt}
\email[]{kliemt@physik.uni-frankfurt.de}
\affiliation{Physikalisches Institut, Goethe-Universit\"at Frankfurt/M, 60438 Frankfurt/M, Germany}
\author{J.Banda}
\affiliation{Max Planck Institute for Chemical Physics of Solids, 01187 Dresden, Germany}
\author{C.Geibel}
\affiliation{Max Planck Institute for Chemical Physics of Solids, 01187 Dresden, Germany}
\author{M.Brando}
\affiliation{Max Planck Institute for Chemical Physics of Solids, 01187 Dresden, Germany}
\author{C.Krellner}
\affiliation{Physikalisches Institut, Goethe-Universit\"at Frankfurt/M, 60438 Frankfurt/M, Germany}
\date{\today}

\begin{abstract}
We present the crystal growth as well as the structural, chemical and physical chracterization of SmRh$_2$Si$_2$ single crystals. Their ground state is antiferromagnetic, as indicated by the behaviour of the magnetic susceptibility and the specific heat at the second order phase transition observed at $T_{\rm N} = 64\,\rm K$. The Sommerfeld coefficient is small and similar to that of LuRh$_2$Si$_2$ with $\gamma_0\approx 7\,\rm mJ/(molK^2)$. Susceptibility measurements show no Curie-Weiss behaviour at high temperatures which is a consequence of the large Van-Vleck contribution of the excited multiplets of Sm$^{3+}$. Previous angle-resolved photoemission studies showed that at $10\,\rm K$, the valence of the Sm ions is smaller than three at the surface as well as in the bulk, suggesting a possible Kondo screening of the Sm$^{3+}$ ions. This could not be observed in our thermodynamic and transport measurements. 

\end{abstract}

% PACS, the Physics and Astronomy Classification Scheme.
%local moments, magnetic anisotropy, metals and alloys, antiferromagnetics
%\keywords{}
\maketitle

\def\neel{{N\'eel} }
\def\FA{F_{\rm an}}
\def\CA{C_{\rm an}}
\def\MS{M_{\rm sat}}
\def\EDF{E_{\rm df}  }
\def\text#1{{\rm #1}}
\def\i{\item}
\def\[{\begin{eqnarray*}}
\def\]{\end{eqnarray*}}
\def\bv{\begin{verbatim}}
\def\ev{\end{verbatim}}
\def\ganz{Z}
\def\3{\ss}
\def\reel{{\cal}R}
\def\platz{\;\;\;\;}
\def\beginvector{\left(\begin{array}{c}   }
\def\endvector{\end{array}\right)}
\def\fff{\frac{3}{k_B} F }
\def\vec#1{ {\rm \bf #1  } }
\def\KBMUEF{\frac{3k_B}{\mu_{\rm eff}^2 }}

%%%%%%%%%%%%%%%%%%%%%%%%%%%%%%%%%%%%%%%%%%%%%%%%%%%%%%%%%%%%%%

\section{Introduction}
Sm-based compounds have been studied rarely in the past compared to compounds containing lanthanide ions like Ce, Eu, Gd, Ho or Yb,  but the recent discovery of topologically insulating behaviour in SmB$_6$ \cite{Takimoto2011, Lu2013, Kim2014} renewed the interest in the community in Sm-based compounds. 
Some Sm compounds exhibit strongly correlated electron behaviour like the
heavy-fermion compounds SmOs$_4$Sb$_{12}$ \cite{Sanada2005} or SmT$_2$Al$_{20}$ \cite{Higashinaka2011}, which  have a complex crystallographic structure and the characterization of systems with lower structural complexity is highly desired.
Recently, the valence of the Sm ions in SmRh$_2$Si$_2$ single crystals was studied by angle resolved photoemission spectroscopy (ARPES). Contributions from the bulk as well as from the surface to the photoemission spectrum have been determined \cite{Chikina2017}. Divalent Sm $4f$ contributions to the photoemission spectrum are expected close to the Fermi level and have been observed experimentally in the past for Sm metal \cite{Wertheim1978} and a number of Sm compounds \cite{Allen1980, Yamaoka2012}. Such emissions from divalent Sm were mostly identified as surface valence transitions in trivalent bulk systems. 
ARPES revealed that for Sm ions in SmRh$_2$Si$_2$, the electronic properties of bulk and surface are rather similar and the Sm ions behave in both cases slightly mixed valent. 
The mean valence of the Sm ions in SmRh$_2$Si$_2$ was estimated to be about $2.94$ at $T=10\,\rm K$ \cite{Chikina2017}.
Usually, the valence of Sm ions at the surface of a single crystal is $2+$ and the observed deviation is quite unusual. In the bulk, small deviations from the $3+$ valence usually occur in Ce or Yb compounds which show the Kondo effect. 
This motivated the study of the bulk properties to see, if these small deviations influence the thermodynamic and transport properties of SmRh$_2$Si$_2$.

The crystallographic ThCr$_2$Si$_2$-type structure of SmRh$_2$Si$_2$, Fig.~\ref{structure} (a), was reported by Felner and Nowik \cite{Felner1983a, Felner1984} and an isothermal section ($T = 870\,\rm K$) of the ternary Sm-Rh-Si phase diagram was explored by Morozkin \textit{et al.} \cite{Morozkin1996, Morozkin1996a}. Susceptibility as well as electrical resistivity measurements on polycrystalline samples were reported by different authors:
Signatures of magnetic transitions in the susceptibility at $62$, $35$ and $10\,\rm K$ and a kink in the resistivity at $60.5\,\rm K$ were observed by Kochetkov \textit{et al.} \cite{Kochetkov1997}. Felner and Nowik reported about peaks in the susceptibility at $8\pm 2\,\rm K$ and $46 \pm 2\,\rm K$ \cite{Felner1984}. These conflicting results in literature, call for a precise determination of the magnetic and thermodynamic properties in high-quality single crystals.

\section{Experiment}

Single crystals of SmRh$_2$Si$_2$ were grown in In flux. 
The high purity starting materials 
Sm (99.99\%, REP), Rh (99.9\%, Heraeus), 
Si (99.9999\%, Wacker), with the molar ratio of 1:2:2,
and In (99.9995\%, Schuckard) were weighed 
in a graphite crucible with a volume of 10 ml and sealed in a niobium crucible 
under argon atmosphere (99.999\%). 
The stoichiometric composition of the elements was used with 96 at.\% indium as flux.
Indium was put at the bottom of the crucible covered by the high-melting elements Rh and Si covered again by indium pieces. Subsequently, the crucible was transfered to the Ar-filled glove-box, where Sm was placed on top, covered once more by indium. Finally, the Nb-crucible was closed using arc-welding.  
The filled Nb-crucible was put under a stream of Ar in a resistive furnace (GERO HTRV 70-250/18) and the elements were heated
up to 1550$^{\circ}$C. The melt was homogenized for 1 hr and then cooled by slow moving the whole furnace with 1 mm/hrs leading to a cooling rate in the range of 0.5 - 4 K/hrs down to 1000$^{\circ}$C, while the position of the crucible stayed fixed. 
The growth parameters have been chosen similarily as reported for the growth of GdRh$_2$Si$_2$ \cite{Kliemt2015}.

Until now, the successful single crystal growth was not reported, which probably is due to the incongruent melting of this material at approximately 1760$^{\circ}$C \cite{Morozkin1997}.  
We therefore employed the flux-growth technique as described above, which allows to perform the crystal growth at temperatures below the high melting points of the elements
(Sm 1072$^{\circ}$C, Rh 1964$^{\circ}$C, and Si 1414$^{\circ}$C). 
So far, it was not possible to determine the accurate crystallization temperature of SmRh$_2$Si$_2$ in 96\, at.\% In, but a systematic optimization of the temperature profile revealed that the largest crystals could be grown when starting the crystal growth at T$_{\rm start}$= 1520$^\circ$C, compared to growths starting below 1500$^\circ$C. 
After cooling, the excess In was dissolved by hydrochloric acid until phase pure SmRh$_2$Si$_2$ was obtained.
Although, not everything of the starting material crystallized in the SmRh$_2$Si$_2$ phase, the resulting single crystals of SmRh$_2$Si$_2$ are large enough to carry out several physical characterization measurements. Typical single crystals as the one shown in inset Fig.~\ref{laue} (c) have a platelet habit with the shortest dimension, $100-400\,\mu$m, along the crystallographic $c$-direction and $1-2\,\rm mm$ perpendicular to that direction. Few single crystals had an extension in $c$-direction of about $1\,\rm mm$ and thus were suitable for transport measurements along this direction.

The chemical composition determined by energy-dispersive 
X-ray spectroscopy (EDX) revealed (20$\pm$1)at.$\%$ Sm,
(40$\pm$1)at.$\%$ Rh and (40$\pm$2)at.$\%$ Si. 
We cannot exclude some Rh-Si site exchange within these error bars; however, for the related compound YbRh$_2$Si$_2$
a detailed structural and chemical analysis was conducted by highly accurate X-ray diffraction and wavelength dispersive X-ray spectroscopy measurements \cite{wirth2012}. There, it turned out that the structure accepts some
Rh-Si site exchange, with a rather small homogeneity range for Rh, which was found to be 40.0 - 40.2\, at.\%.

The crystal structure was characterized by powder 
X-ray diffraction (PXRD) on crushed single crystals, using 
Cu-K$_{\alpha}$ radiation.
PXRD confirmed the $I4/mmm$ tetragonal structure 
with lattice parameters 
a = 4.0555(2) \AA\,  and c =10.0402(4) \AA, which is in agreement with the 
data published for polycrystalline samples \cite{Felner1983a}. 

The orientation of the single crystals was determined 
using a Laue camera with X-ray radiation from a 
tungsten anode. 
The high quality of the single crystals is evident also from a Laue back scattering image, presented in Fig.~\ref{laue} (b). The central point can be indexed as the (001) reflex, proving that the direction perpendicular to the surface of the platelets corresponds to the $c$-direction. 
Four-point resistivity, magnetization, and heat-capacity measurements were performed
using the commercial measurements options of a Quantum Design Physical Properties Measurement System.

\section{Results and discussion}

\subsection*{Heat capacity and entropy}
In Fig.~\ref{SmRh2Si2_HC} (a) specific-heat data are shown for the two related materials SmRh$_2$Si$_2$ and LuRh$_2$Si$_2$. The latter serves as a non-magnetic reference system with a completely filled $4f-$shell. The LuRh$_2$Si$_2$ data were taken from Ref.~\cite{Ferstl2007}. For SmRh$_2$Si$_2$ a pronounced and sharp $\lambda$-type anomaly is observed at  $T_{\rm N}=64\,\rm K$, establishing a second order phase transition into the antiferromagnetically (AFM) ordered phase. 
Inset (b) in Fig.~\ref{SmRh2Si2_HC} displays $C/T$ versus $T^2$. Between $20\,{\rm K}^2 < T^2 < 60\,{\rm K}^2$, $C/T$ of SmRh$_2$Si$_2$ is nearly equal to $C/T$ of LuRh$_2$Si$_2$ which shows that magnetic excitations are frozen out indicating the presence of a gap in the magnon spectra. Above $T^2 = 60\,{\rm K}^2$, $C/T$ of SmRh$_2$Si$_2$ increases much stronger than $C/T$ of LuRh$_2$Si$_2$ which exhibits a nearly linear behavior. This much stronger increase corresponds to the exponential thermal activation of magnetic excitations across the gap. This gap very likely results from the magnetic anisotropy not only between the $c$ and the in-plane direction, but also within the plane as evidenced by magnetization data (see below discussion of Fig. 4). While both LuRh$_2$Si$_2$ and LaRh$_2$Si$_2$ show a linear behavior in $C/T(T)$ versus $T^2$ for $T^2 < 60\,{\rm K}^2$, SmRh$_2$Si$_2$ still present a small positive curvature resulting in slightly larger values of $C/T$ than in LuRh$_2$Si$_2$ at lowest $T$. This indicate the onset of an additional contribution to $C/T$ at lowest $T$, which is likely connected with the nuclear contribution, but might also origin from magnetic defects. The presence of this additional low $T$ contribution prevents a direct determination of the Sommerfeld coefficient $\gamma_0$ and of the low $T$ Debye temperature $\Theta_D$. However, the observation that for $T^2 < 80\,\rm K^2$ SmRh$_2$Si$_2$ presents nearly the same $C/T$ as LuRh$_2$Si$_2$ indicate that $\gamma_0$ and $\Theta_D$ of SmRh$_2$Si$_2$ are very close to those of LuRh$_2$Si$_2$, $\gamma_0\approx 7 \rm mJ/(molK^2)$ and $\Theta_D = 380\,\rm K$.
 %nuclear contribution in the order of $\approx 1\rm mJ/molK^2$ \cite{Takeda2008} 

\begin{figure}
\centering
\includegraphics[width=1.0\columnwidth]{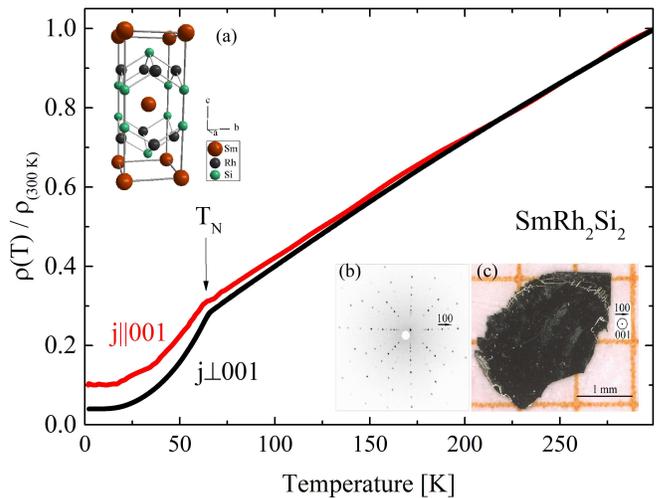}
\caption{Temperature dependence of the normalized resistivity $\rho(T)/\rho(300\,\rm K)$ for $j\parallel c$ and $j\perp c$. At T$_{\rm N}=65\,\rm K$ the onset of antiferromagnetic order is indicated by kinks in the curves for both current directions. {\it Inset:} (a) Tetragonal ThCr$_2$Si$_2$-type structure; (b) The Laue pattern taken along the 001-direction shows the 4-fold symmetry.
The sharp reflexes are indicators for the good sample quality; (c) Optical microscope image of a SmRh$_2$Si$_2$ single crystal.} 
\label{structure}\label{laue}\label{SmRh2Si2_Rho}
\end{figure}
\begin{figure}
\centering
\includegraphics[width=1.0\columnwidth]{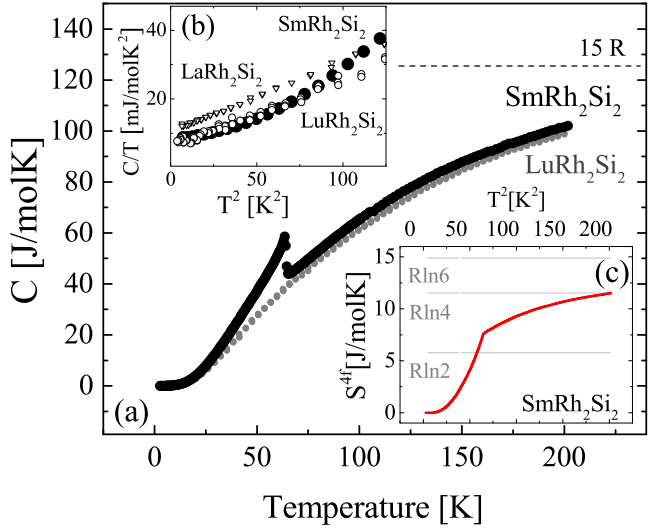}
\caption{(a) Specific-heat data as function of temperature for a single crystal of SmRh$_2$Si$_2$ and polycrystalline LuRh$_2$Si$_2$ (Ref.~\cite{Ferstl2007}). The high-temperature Dulong-Petit-value of $15\,\rm R$ is not reached at $200\,\rm K$. Inset (b) enlarges the low-temperature part of the specific heat of SmRh$_2$Si$_2$ (black circles), LuRh$_2$Si$_2$ (open circles) and LaRh$_2$Si$_2$ (open triangles, Ref.~\cite{Berisso2002}) plotted as $C/T$ versus $T^2$. Inset (c) shows the magnetic entropy.} 
\label{SmRh2Si2_HC}
\end{figure}

\begin{figure}
\centering
\includegraphics[width=1.0\columnwidth]{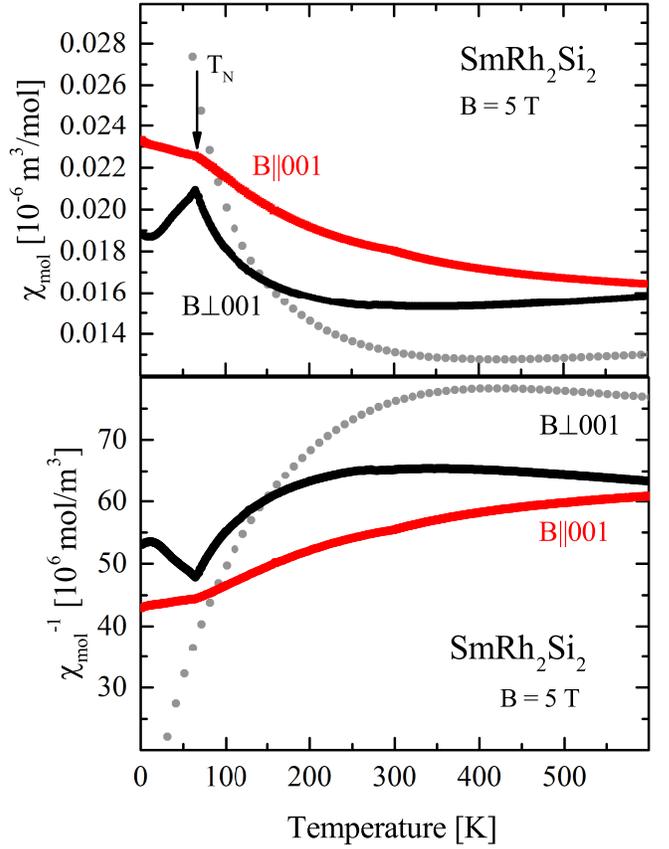}
\caption[SmRh$_2$Si$_2$: Susceptibility $\chi(T)$ and inverse susceptibility $\chi^{-1}(T)$]{SmRh$_2$Si$_2$ {\it Top:} Susceptibility $\chi(T)$ as a function of temperature for $B\parallel 001$ and $B\perp 001$. The susceptibility was calculated using the standard expression including Curie and Van-Vleck contribution and with the first excited $J=7/2$ multiplet set at $1500\,\rm K$ (grey dots). {\it Bottom:} Because of the dominant $T$-independent Van-Vleck contributions, a tendency towards a Curie-Weiss behaviour in the inverse susceptibility is only seen at low temperatures, just above $T_{\rm N}$. The inverse susceptibility $ \chi^{-1}(T)$ does not show Curie-Weiss behaviour at high temperatures. }
\label{SmRh2Si2_ChivT}
\end{figure}

By subtracting the heat capacity of the non-magnetic reference compound LuRh$_2$Si$_2$ \cite{Ferstl2007} we obtained the magnetic part of the specific heat $C^{\rm 4f}$ and by integrating $C^{\rm 4f}/T$ the entropy $S^{\rm 4f}$. Fig.~\ref{SmRh2Si2_HC} inset (c)
 shows that $S^{\rm 4f}$ increases up to the phase transition and increases further towards higher temperature.
The Sm ion with $J=5/2$ has 6 possible states and 
the entropy of the full crystalline electric field (CEF) multiplet would amount to $S^{\rm 4f}= R\,\rm ln\,6$ at the phase transition.
We find that the entropy is strongly influenced by the CEF. It is smaller than $R\,\rm ln\,6$ and larger than $R\,\rm ln\,2$ at the phase transition which shows that besides the ground state doublet also the occupation of higher levels contributes to the entropy.

\subsection*{Electrical resistivity}
Electrical transport data shown in Fig.~\ref{SmRh2Si2_Rho}, indicate an anisotropic behaviour 
for current flowing parallel and perpendicular to the $[001]$-direction below $T_{\rm N}$. 
We determined $RR^{j\parallel c}_{1.8\,\rm K} = 10$ and $RR^{j\perp c}_{1.8\,\rm K} = 25$.
The absolute value of $\rho(T)$ at room temperature for the in-plane resistivity was about $80\,\mu\Omega\rm cm$. For better comparison of the data with $j\parallel c$ and $j\perp c$, we present the normalized resistivity $\rho(T)/\rho_{300\,\rm K}$.
For both directions, the resistivity shows a quasi-linear-in-$T$ behaviour from $300\,\rm K$ to the AFM phase transition. 
Upon cooling, at $T_{\rm N}$ a change of the slope of the resistivity curves occurs and the decrease of the resistivity becomes stronger. 
In previous work, $\rho_{200K}/\rho_{0}\sim 5$ was determined for polycrystalline material \cite{Kochetkov1997}.
The residual resistivity ratios $\rm RRR=\rho_{300K}/\rho_{0}\sim 25$ (corresponding to $\rho_{200K}/\rho_{0}\sim 17.5$) determined on our samples for $j\perp c$ shows that we succeeded in growing high-quality crystals using the indium-flux method. 

\subsection*{Magnetization}
The magnetic moment of the Sm ions in SmRh$_2$Si$_2$ is small and great care needs to be taken to subtract the background contribution. We therefore studied this material by using large single crystals with a mass of about $30\,\rm mg$. 
In Fig.~\ref{SmRh2Si2_ChivT}, the temperature dependence of the susceptibility for a field of $B=5\,\rm T$ along two main symmetry directions is shown.
Upon cooling, the transition into the AFM-ordered state appears at the \neel temperature $T_{\rm N}=65\,\rm K$. For $B\perp c$, the susceptibility decreases strongly below  $T_{\rm N}$ and shows a small increase at lowest temperatures. This feature varies from sample to sample and is probably caused by a paramagnetic defect contribution. The comparison of the susceptibility curves for both field directions suggests that below $T_{\rm N}$ the magnetic moments are ordered within the basal plane of the tetragonal lattice. Additionally, our data shows the absence of further magnetic transitions which were proposed by Ref.~\cite{Felner1983a} and Ref.~\cite{Kochetkov1997} in previous works.
Above $T_{\rm N}$, the $T$ dependence of the susceptibility is typical for a Sm$^{3+}$ system, cf. Fig.~\ref{SmRh2Si2_ChivT}. 
It first shows in a small $T$ range a Curie-Weiss like decrease, then evolves towards a large $T$ independent susceptibility, 
and for field along the $c$ axis even show a slight increase towards higher temperatures. 
This peculiar $T$ dependence which significantly differs from the Curie-Weiss behaviour 
observed for most other rare earth can easily be accounted for by the specific properties 
of the $4f$ state of Sm$^{3+}$. In the $J = L-S = 5/2$ ground state multiplet of the $4f^{5}$ state of Sm$^{3+}$ 
the almost cancellation of the orbital part of the moment with $L = 5$, $g_{\rm orb} = 1$ by the spin part with 
$S = 5/2$, $g_{\rm spi} = 2$ results in a tiny Land\'{e} factor $g_L = 0.286$. 
This leads to a quite small value of the effective moment, $\mu_{\rm eff} = 0.84 \mu_{\rm B}$ and, 
accordingly, to a small Curie-Weiss contribution. In contrast, the first excited multiplet $J = 7/2$ 
being at a comparatively low energy $E_{7/2}\approx 1500\,\rm K$ \cite{Osborn1991} results in a large $T$-independent 
Van-Vleck type susceptibility which overwhelms the Curie contribution at $T > 200\,\rm K$. 
Further on the increase of the susceptibility for $T > 400\,\rm K$ is due to thermal excitation of the 
$J = 7/2$ multiplet, because it has a much larger Land\'{e} factor $g_L = 0.825$, 
and thus a much larger effective moment $\mu_{\rm eff} = 3.28 \mu_{\rm B}$. 
Therefore the prefactor of its Curie contribution is a factor of 15 larger than that of the $J = 5/2$
 ground state multiplet. The further excited $J$ multiplets enhance this behaviour. 
The observed susceptibility, cf. Fig.~\ref{SmRh2Si2_ChivT}, qualitatively agrees with that calculated for an isolated Sm$^{3+}$ ion using the standard expression for Sm$^{3+}$ and Eu$^{3+}$ systems, see e.g. Ref.~\cite{Takikawa2010} 
(Fig.~\ref{SmRh2Si2_ChivT}, grey dots), 
but presents a number of differences, e.g a clear anisotropy. 
This anisotropy is likely dominated by the effect of the crystal electric field which affects 
both, the Curie contribution of the different multiplet states as well as the Van-Vleck contribution. 
But especially at low temperatures it is also affected by the exchange interaction, which also seems to be anisotropic. 
Due to the low susceptibility of the Sm$^{3+}$ ions, contributions of the conduction electrons namely Pauli, orbital and diamagnetic susceptibility need to be taken into account.
At high temperatures, the difference between the calculated curve (grey dots) 
and the measured susceptibility is about $2.0\cdot 10^{-9} {\rm m}^3/{\rm mol}$.
This is in perfect agreement with the susceptibility 
of the non-$4f$ electrons determined for LaRh$_2$Si$_2$ by Ref.~\cite{Sekizawa1987}.
%%%%%%%%%%%%%%%%%%%

\begin{figure}
\centering
\includegraphics[width=1.0\columnwidth]{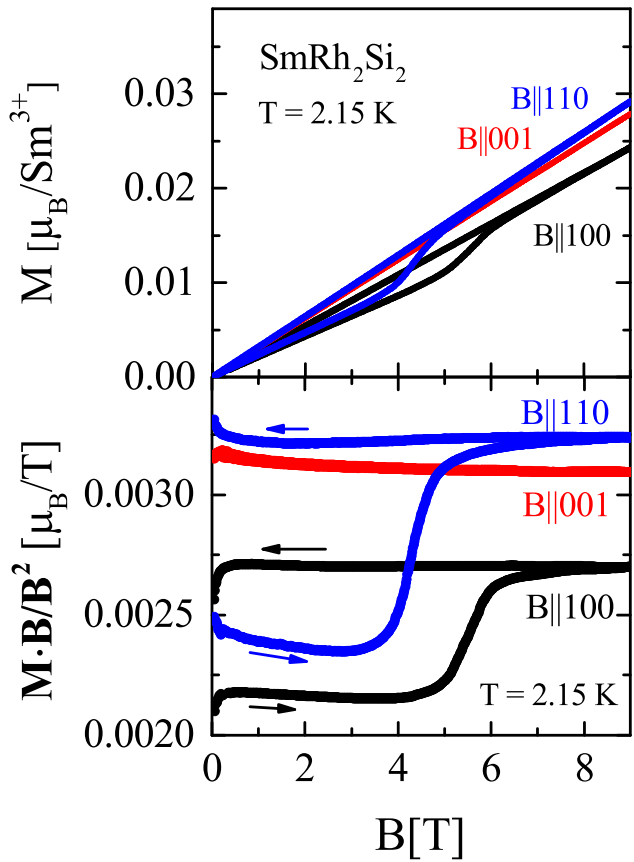}
\caption[SmRh$_2$Si$_2$: Magnetization $M(B)$ at $T=2.15\,\rm K$]{SmRh$_2$Si$_2$ {\it Top:} Field dependence of the magnetization $M=\vec{M}\cdot\vec{B}/B$ at $T=2.15\,\rm K$ for field applied along the three main symmetry directions. {\it Bottom:} The figure shows the data of the upper panel plotted as $\vec{M}\cdot\vec{B}/B^2$.}
\label{SmRh2Si2_MvH}
\end{figure}

The field dependence of the magnetization was investigated 
at $T=2.15\,\rm K$ in fields up to $B=9\,\rm T$. In the upper panel of Fig.~\ref{SmRh2Si2_MvH}, the magnetization for fields applied along the main symmetry directions is shown. Here, the magnetization shows anisotropic hystereses effects. 
While the slope of $M$ is nearly constant for field $B$ along the $[001]$ direction, red curve in Fig.~\ref{SmRh2Si2_MvH}, for $B\perp 001$, it is smaller and shows hysteresis effects, black and blue curves. This field dependent behaviour is consistent with a moment orientation in the basal plane of the tetragonal lattice. 
The occurence and size of the hystereses effects is strongly sample dependent.

\section{Summary}

We have grown SmRh$_2$Si$_2$ single crystals for the first time which enables us not only to determine precisely the phase transition temperature but also to study the anisotropic behaviour of this compound via electrical transport and magnetization measurements.
In this work, single crystals of SmRh$_2$Si$_2$ have been grown by a modified Bridgman method from indium flux. After an optimization of the temperature-time profile of the growth experiment, we obtained millimetre-sized single crystals with a platelet habit with the $c$ axis perpendicular to the platelet. PXRD measurements on crushed single crystals
confirmed the $I4/mmm$ tetragonal structure. The lattice parameters 
are in agreement with the data published for polycrystalline samples.
The specific heat of SmRh$_2$Si$_2$ shows only one sharp $\lambda$-type anomaly at $T_{\rm N}=64\,\rm K$, establishing a second order phase transition into the AFM ordered phase. 
Magnetic measurements on the single crystals show the ordering of the Sm$^{3+}$ moments at $T_{\rm N}$ as well as the reorientation of magnetic domains for an external field applied in the basal plane of the tetragonal lattice.
Above $T_{\rm N}$, the susceptibility presents the typical behavior of a Sm$^{3+}$ system. 
It shows a small Curie-Weiss contribution below $100\,\rm K$ due to the tiny Land\'{e} factor of the $J = 5/2$ ground state 
multiplet, but a large $T$-independent Van-Vleck contribution which dominates the susceptibility above $200\,\rm K$ 
and results from low lying excited $J$ multiplets of Sm$^{3+}$.
Below $T_{\rm N}$, electrical transport data show an anisotropy between the resistivity measured with current parallel and perpendicular to the $c$ axis. The residual resistivity ratio $\rm RRR=\rho_{300K}/\rho_{0}\sim 25$ indicates that we succeeded in growing high-quality single crystals from a high-temperature indium flux.
Recently, the single crystals enabled the investigation of their bulk and surface properties by ARPES. 
It turned out that at $T=10\,\rm K$ the valence of the Sm ions is smaller than three at the surface as well as in the bulk \cite{Chikina2017}. This hints to a possible Kondo screening of the Sm$^{3+}$ ions but we could not detect any effect in the transport and thermodynamic properties.

%%%%%%%%%%%%%%%%%%%%%%%%%%%%%%%%%%%%%%%%%%%%%%%%%%%%%%%%%%%%%%%%%%%%%%%%%%%%%%%%%%%%%%%%%%%%%

\begin{acknowledgments}
We are indebted to B. Schmidt for useful discussions.
 KK, CK and MB grateful acknowledge support by the German Research Foundation (DFG) by Grants No. KR 3831/4-1 and No. BR 4110/1-1.
\end{acknowledgments}

%%%%%%%%%%%%%%%%%%%%%%%%%%%%%%%%%%%%%%%%%%%%%%%%%%%%%%%%%%%%%%%%%%%%%%%%%%%%%%%%%

%%%%%%%%%%%%%%%%%%%%%%%%%%%%%%%%%%%%%%%%%%%%%%%%%%%%%%%%%%%%%%%%%%%%%%%%%%%%%%%%%%%%%

%%%%%%%%%%%%%%%%%%%%%%%%%%%%%%%%%%%%%%%%%%%%%%%%%%%%%%%%%%%%%%%%%%%%%%%%%%%%

%\begin{thebibliography}{}
\bibliographystyle{unsrt}
\bibliography{SmRh2Si2_PRB}

\begin{thebibliography}{10}

\bibitem{Takimoto2011}
T.~Takimoto.
\newblock {\em J.Phys.Soc.Jpn.}, 80:123710, 2011.

\bibitem{Lu2013}
F.~Lu, J.~Zhao, H.~Weng, Z.~Fang, and X.~Dai.
\newblock {\em Phys.Rev.Lett.}, 110:096401, 2013.

\bibitem{Kim2014}
D.~J. Kim, J.~Xia, and Z.~Fisk.
\newblock {\em Nat.Mater.}, 13:466, 2014.

\bibitem{Sanada2005}
S.~Sanada, Y.~Aoki, H.~Aoki, A.~Tsuchiya, D.~Kikuchi, H.~Sugawara, and H.~Sato.
\newblock {\em J.Phys.Soc.Jpn.}, 74:246, 2005.

\bibitem{Higashinaka2011}
R.~Higashinaka, T.~Maruyama, A.~Nakama, R.~Miyazaki, Y.~Aoki, and H.~Sato.
\newblock {\em J.Phys.Soc.Jpn.}, 80:093703, 2011.

\bibitem{Chikina2017}
A.~Chikina, A.~Generalov, K.~Kummer, M.~G\"uttler, V.~N. Antonov, Yu.
  Kucherenko, K.~Kliemt, C.~Krellner, S.~Danzenb\"acher, T.~Kim, P.~Dudin,
  C.~Geibel, C.~Laubschat, and D.~V. Vyalikh.
\newblock {\em Phys.Rev.B}, 95:155127, 2017.

\bibitem{Wertheim1978}
G.~K. Wertheim and G.~Crecelius.
\newblock {\em Phys.Rev.Lett.}, 40:813, 1978.

\bibitem{Allen1980}
J.~W. Allen, L.~I. Johansson, I.~Lindau, and S.~B. Hagstrom.
\newblock {\em Phys.Rev.B}, 21:1335, 1980.

\bibitem{Yamaoka2012}
H.~Yamaoka, P.~Thunstr\"om, I.~Jarrige, K.~Shimada, N.~Tsujii, M.~Arita,
  H.~Iwasawa, H.~Hayashi, J.~Jiang, T.~Habuchi, D.~Hirayama, H.~Namatame,
  M.~Taniguchi, U.~Murao, S.~Hosoya, A.~Tamaki, and H.~Kitazawa.
\newblock {\em Phys.Rev.B}, 85:115120, 2012.

\bibitem{Felner1983a}
I.~Felner and I.~Nowik.
\newblock {\em Solid State Commun.}, 47:831, 1983.

\bibitem{Felner1984}
I.~Felner and I.~Nowik.
\newblock {\em J.Phys.Chem.Solids}, 45:419, 1984.

\bibitem{Morozkin1996}
A.~V. Morozkin, Yu.~D. Seropegin, and O.~I. Bodak.
\newblock {\em J.Alloy.Compd.}, 234:143, 1996.

\bibitem{Morozkin1996a}
A.~V. Morozkin and Yu.~D. Seropegin.
\newblock {\em J.Alloy.Compd.}, 237:124, 1996.

\bibitem{Kochetkov1997}
Yu.~V. Kochetkov, V.~N. Nikiforov, S.~A. Klestov, and A.~V. Morozkin.
\newblock {\em J.Magn.Magn.Mater.}, 157:665, 1997.

\bibitem{Kliemt2015}
K.~Kliemt and C.~Krellner.
\newblock {\em J. Cryst. Growth}, 419:37, 2015.

\bibitem{Morozkin1997}
A.~V. Morozkin, Yu.~D. Seropegin, A.~V. Gribanov, and J.~M. Barakatova.
\newblock {\em J.Alloy.Compd.}, 256:175, 1997.

\bibitem{wirth2012}
S.~Wirth, S.~Ernst, R.~Cardoso-Gil, H.~Borrmann, S.~Seiro, C.~Krellner,
  C.~Geibel, S.~Kirchner, U.~Burkhardt, Y.~Grin, and F.~Steglich.
\newblock {\em J.Phys.Condens.Matter}, 24:294203, 2012.

\bibitem{Ferstl2007}
J.~Ferstl.
\newblock {\em New Yb-based systems: From an intermediate-valent to a
  magnetically ordered state}.
\newblock PhD thesis, TU Dresden, Cuvillier G\"ottingen, 2007.

\bibitem{Berisso2002}
M.~G{\'o}mez~Berisso, P.~Pedrazzini, J.G. Sereni, O.~Trovarelli, C.~Geibel, and
  F.~Steglich.
\newblock {\em The European Physical Journal B - Condensed Matter and Complex
  Systems}, 30:343, 2002.

\bibitem{Osborn1991}
R.~Osborn, S.W. Lovesey, A.D. Taylor, and E.~Balcar.
\newblock {\em {Handbook on the Physics and Chemistry of Rare earths}},
  volume~14.
\newblock Elsevier, North Holland, 1991.

\bibitem{Takikawa2010}
Y.~{Takikawa}, S.~{Ebisu}, and S.~{Nagata}.
\newblock {\em Journal of Physics and Chemistry of Solids}, 71:1592, 2010.

\bibitem{Sekizawa1987}
K.~Sekizawa, Y.~Takano, H.~Takigami, and Y.~Takahashi.
\newblock {\em J. Less Common Met.}, 127:99, 1987.

\end{thebibliography}
%\bibliography{../Bibtex/diss}          %%%einkommentieren!!
%\end{thebibliography}

%\bibliography{}

\end{document}